\documentclass[reprint,amsmath,amssymb,aps,prl,superscriptaddress,longbibliography]{revtex4-2}

\usepackage{graphicx}% Include figure files
\usepackage{soul}
\usepackage[usenames]{xcolor}
\usepackage{dcolumn}% Align table columns on decimal point
\usepackage{bm}% bold math
\newcommand{\TMCT}{T_\mathrm{MCT}}
\newcommand{\giso}{\gamma_\mathrm{iso}}
\newcommand{\gdev}{\gamma_\mathrm{dev}}
\newcommand{\gext}{\gamma_\mathrm{ext}}

\newcommand{\eext}{\bm{e}_\mathrm{ext}}
\newcommand{\ecom}{\bm{e}_\mathrm{com}}
\newcommand{\ghop}{g_\mathrm{hop}}
\newcommand{\tghop}{\tilde{g}_\mathrm{hop}}
\newcommand{\RestrictedTo}[2]{\left.\kern-\nulldelimiterspace #1 \vphantom{\big|} \right|_{#2}} % from https://tex.stackexchange.com/a/22255

\begin{document}

\title{Elastoplasticity Mediates Dynamical Heterogeneity Below the Mode-Coupling Temperature}

\author{Rahul N. Chacko}
\affiliation{Department of Physics and Astronomy, University of Pennsylvania, Philadelphia, Pennsylvania 19104, USA}
\affiliation{James Franck Institute, University of Chicago, Chicago, Illinois 60637, USA}

\author{Fran\c{c}ois P. Landes}
\affiliation{Universit\'{e} Paris-Saclay, CNRS, Laboratoire Interdisciplinaire des Sciences du Num\'{e}rique, 91400, Orsay,
France}

\author{Giulio Biroli}
\affiliation{Laboratoire de Physique de l'\'{E}cole Normale Sup\'{e}rieure, ENS, Universit\'{e} PSL, CNRS, Sorbonne Universit\'{e}, Universit\'{e} Paris-Diderot, Sorbonne Paris Cit\'{e}, 75005 Paris, France}

\author{Olivier Dauchot}
\affiliation{UMR Gulliver 7083 CNRS, ESPCI, PSL Research University, 10 rue Vauquelin, 75005 Paris, France}

\author{Andrea J. Liu}
\affiliation{Department of Physics and Astronomy, University of Pennsylvania, Philadelphia, Pennsylvania 19104, USA}

\author{David R. Reichman}
\affiliation{Department of Chemistry, Columbia University, 3000 Broadway, New York, New York 10027, USA}

\date{\today}

\begin{abstract}
As liquids approach the glass transition temperature, dynamical heterogeneity emerges as a crucial universal feature of their behavior. Dynamic facilitation, where local motion triggers further motion nearby, plays a major role in this phenomenon. Here we show that long-range, elastically-mediated facilitation appears below the mode-coupling temperature, adding to the short-range component present at all temperatures. Our results suggest deep connections between the supercooled liquid and glass states, and pave the way for a deeper understanding of dynamical heterogeneity in glassy systems.
\end{abstract}

\maketitle

Dynamic facilitation is the process whereby relaxation of a local region in a glassy system enables another local region to subsequently move and relax.
Facilitation plays a major role in the spatiotemporal pattern of correlated relaxation events characteristic of supercooled liquids, referred to as ``dynamical heterogeneity," which is perhaps the most striking hallmark of glassy dynamics~\cite{berthier2011dynamical}. The underlying structural and  microscopic origins of dynamic facilitation (also called kinetic facilitation) and dynamical heterogeneity are still widely debated. At high temperatures 
facilitation appears to be a local, short-ranged process~\cite{Fredrickson1984,Chandler2010,Candelier2010,Keys2011,Elmatad2012}.  However it has been appreciated for decades that some relaxation processes in glassy media involve long-ranged correlations. Given that dynamic facilitation in supercooled liquids arises initially at short time scales, where the system should behave effectively as a solid, it is natural to wonder whether long-ranged elastic processes may influence facilitation and dynamical heterogeneity.
   
An extreme example of the influence of elasticity is the low-temperature relaxation of tunneling defects in glasses,
which is mediated by long-ranged (dipolar) phonon exchange~\cite{Anderson1972,Phillips1972,Joffrin1975}.
Another example is provided by the plastic behavior of amorphous solids.
Local plastic deformations in elastic media induce signature quadrupolar perturbations to the stress and strain fields in the surrounding material which decay as power laws in space~\cite{Picard2004}.
The role of such fields in triggering relaxation elsewhere when an amorphous solid is mechanically deformed has been the focus of intense research activity during the last decade~\cite{Martens2011,Dasgupta2012,Gendelman2015,Patinet2016,Budrikis2017,Nicolas2018,Barbot2020}. This triggering by stress is
 a type of ``elastically-mediated'' dynamic facilitation that has been thoroughly  studied in the context of the rheology of amorphous solids by means of elastoplastic models~\cite{Nicolas2018},
in which each rearrangement perturbs the stress of the surrounding material, thereby triggering new rearrangements when the perturbing stress surpasses some local threshold value. It should be noted, however, that elastoplastic models do not incorporate the type of thermal motion that occurs in equilibrium liquids.
Recent reports of anisotropic spatial correlations in stress or strain in liquids~\cite{Levashov2011,Jensen2014,Lemaitre2014,Lemaitre2015,Wu2015,Illing2016} have suggested that
elasticity might play a role even at high temperatures, but it is now understood that such stress correlations must arise for isotropic systems in mechanical equilibrium and do not require
elasticity~\cite{Lemaitre2017,Lemaitre2018}.  

Here we study by numerical simulation the effects of elasticity on dynamic facilitation in supercooled liquids at temperatures above and below the mode-coupling temperature, $\TMCT$. We examine both the strain response to a rearranging particle as well as the pair correlation function of rearrangements, $\ghop (\bm{r})$, characterizing the probability of finding a rearrangement at a relative position $\bm{r}$ within a short time interval following a rearrangement at the origin. 
At all temperatures studied we find that
the strain response of the neighborhood of a rearranging particle is anisotropic and
qualitatively consistent with the response of a linearly elastic medium. 
The pair correlation function of rearrangements, however, reveals a change of behavior with temperature. At high temperatures, dynamic facilitation is {\it local}, \emph{i.e.}~it acts only on nearby regions, as indeed assumed in simple models \cite{Chandler2010,Garrahan2011}. However, 
dynamic facilitation becomes progressively {\it longer-ranged} and {\it elastically-mediated} with decreasing temperature. Remarkably, the emergence of long-ranged facilitation takes place near $\TMCT$, strengthening the interpretation of the mode-coupling crossover as the temperature at which the system starts to display solid-like behaviors~\cite{Biroli2012,Barbot2020,Ozawa2020}.

We conduct NVE molecular dynamics simulations,
using LAMMPS~\cite{LAMMPS},
of a ${d=2}$ dimensional polydisperse system of particles
first introduced in \cite{Berthier2019}.
The swap Monte Carlo algorithm~\cite{Berthier2016,Ninarello2017},
allows for the equilibration of this system
even well below the mode-coupling theory temperature $\TMCT$~\cite{Berthier2019}.
The particles have random diameters $\sigma \in \left[ \sigma_\mathrm{min}, \sigma_\mathrm{max} \right]$
with probability density function $\propto \sigma^{-3}$,
where $\sigma_\mathrm{min}$ and $\sigma_\mathrm{max}$ are determined by the mean $\bar \sigma$ and coefficient of variation $c_\sigma$
of $\sigma$.
Particles interact via a pair potential $V \left( \tilde{r} \right) = G \Theta( \tilde{r}_0 - \tilde{r} )
\left[ \tilde r^{-12} + c_0 + c_2 \tilde r^2 + c_4 \tilde r^4 \right]$.
Here, $\Theta$ is the Heaviside step function and
$c_0$, $c_2$, and $c_4$ are chosen such that $V \left( \tilde r_0 \right) = 0$ and $V^\prime \left( \tilde r \right)$ is continuously differentiable.
The argument $\tilde r = r / \tilde \sigma$
is the interparticle separation $r$ between particles $i$ and $j$ with diameters $\sigma_i$ and $\sigma_j$ normalized
by $\tilde \sigma = \frac{1}{2} \left( \sigma_i + \sigma_j \right) \left( 1 - \epsilon \left| \sigma_i - \sigma_j \right| \right)$.
Following \cite{Berthier2019}, we choose $\tilde r_0 = 2.5$, $\epsilon = 0.2$, and $c_\sigma = 0.23$.
We set $G$, $\bar \sigma$, the (uniform) particle mass $m$, and the Boltzmann constant $k_\mathrm{B}$ to unity,
thus defining our energy, length, mass, and temperature units.
We study $N=10^4$ particles in an $L \times L$ periodic square box of side length $L=100$.
For this system the onset temperature is $T_0 \approx 0.236$ and the mode-coupling temperature is $\TMCT \approx 0.116$ \cite{Berthier2019}.

We study rearrangements in 400 trajectories
with independent initial velocities chosen from the Boltzmann distribution for each of 202 separate equilibrium configurations, 
for eight temperatures $T = 0.100$, $0.105$, $0.110$, $0.115$, $0.120$, $0.130$, $0.140$, and $0.150$,
doubling our statistics by averaging over trajectories in both directions in time~\cite{SM}.
We identify rearranging particles by focusing on quenched trajectories between inherent structures, and by measuring the squared displacement $\Delta r^2$ between
inherent structures~\cite{Lemaitre2014,SM}. 
Our criterion for rearrangement is $\Delta r^2> 5 a^2$ across a time interval $\Delta t = 10^2$,
where $a$ is the plateau height of the root mean squared displacement for particles of similar size in the unquenched system at the given temperature~\cite{SM}.
The time scale $\Delta t =10^2$ matches the time at which dynamic facilitation is at play, identified as the one at which individual rearrangements cluster together in space~\cite{SM}. 
For reference, the ballistic time is ${\tau_\mathrm{ballistic} \approx 10^0}$ in these units, and the relaxation time is $\tau_\alpha = 3 \times 10^2$ and $\tau_\alpha = 2 \times 10^6$ at the highest and lowest temperatures, respectively.
For further characterization and details on the choice of $\Delta t$,
see the Supplementary Material~\cite{SM}.

\begin{figure}[t]
\includegraphics[width=0.48\textwidth]{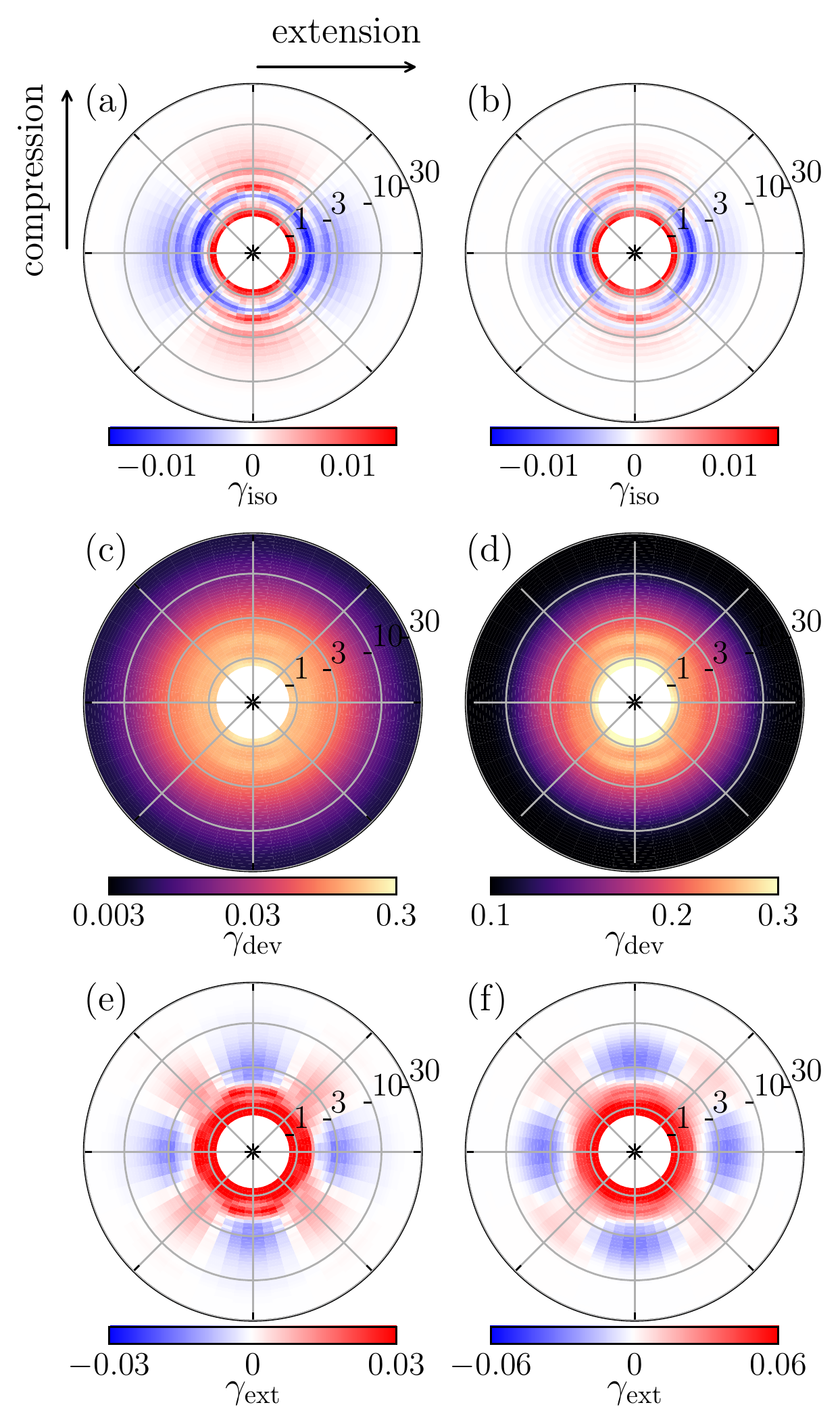}
\caption{Polar plots of
(a,b) $\giso = \mathrm{Tr} \bm{E}$,
(c,d) $\gdev = \sqrt{2 \mathrm{Tr} \bm{E}^{\prime 2}}$, and
(e,f) $\gext = \bm{E}^\prime : \eext \otimes \eext$
across a rearrangement interval $\Delta t = 10^2$
as a function of the position $\bm{r}$ relative to a rearranging particle that was at $\bm{r}=\bm{0}$
at the start of the rearrangement interval
for systems at temperatures $T=0.100$ (a,c,e) and $T=0.150$ (b,d,f).
Here, $\bm{E}$ is the best-fit linear strain tensor at $\bm{r}$,
$\bm{E}^\prime$ is detraced $\bm{E}$,
and $\eext$ is a unit eigenvector corresponding to the largest eigenvalue of $\bm{E}$ at the origin.
The radial axes are plotted on a log scale.
\label{fig:strain_polar}}
\vspace{-3mm}
\end{figure}

Our first observation is that the rearrangement of a given particle causes elastic-like displacements of other particles; \emph{i.e.}~the system, despite being a liquid, displays a solid-like response at short times. To show this, we proceed as follows. 
For each rearrangement in the system, we obtain the local best-fit linear strain tensor $\bm{E}=\bm{E}(\bm{r})$ for all particles from the symmetrized local quenched displacement gradient tensor~\cite{Falk1998,Jensen2014,SM}.
We then extract strain invariants, such as the isotropic strain $\giso = \mathrm{Tr} \bm{E}$ and the deviatoric strain $\gdev = \sqrt{2 \mathrm{Tr} \bm{E}^{\prime 2}}$ (where $\bm{E}^\prime \equiv \bm{E} - \frac{1}{d} \bm{I} \, \mathrm{Tr} \bm{E} $).
Crucially, $\bm{E}$ also gives us local extensional and compressional axes from its orthonormal eigenvectors:
we denote as $\eext$ ($\ecom$) the eigenvector corresponding to the largest (smallest) eigenvalue of $\bm{E}$.
We then average over neighborhoods of rearranging particles translated such that the rearranging particle is at the origin at the start of the rearrangement interval (of duration $\Delta t$) and rotated such that the extensional axis of its local strain tensor is $\eext = \left( 1, 0 \right)$. We also calculate the strain in this direction, $\gext \equiv \bm{E}^\prime : \eext \otimes \eext$, 
where $\eext$ and $\ecom$ are eigenvectors of the local strain tensor $\bm{E}$ computed at the rearranging particle $i$,
whereas $\bm{E}^\prime$ is the local deviatoric strain tensor of the neighbors $j$ of $i$.
The dipolarity of $\giso$, quadrupolarity of $\gext$, and the isotropic behavior of $\gdev$,
shown in Fig.~\ref{fig:strain_polar} for temperatures $T=0.100$ (left) and $T=0.150$ (right)
spanning the full temperature range of our study,
are the expected signatures of the response of an elastic solid~\cite{Zhang2020}.

\begin{figure}[t]
\includegraphics[width=0.48\textwidth]{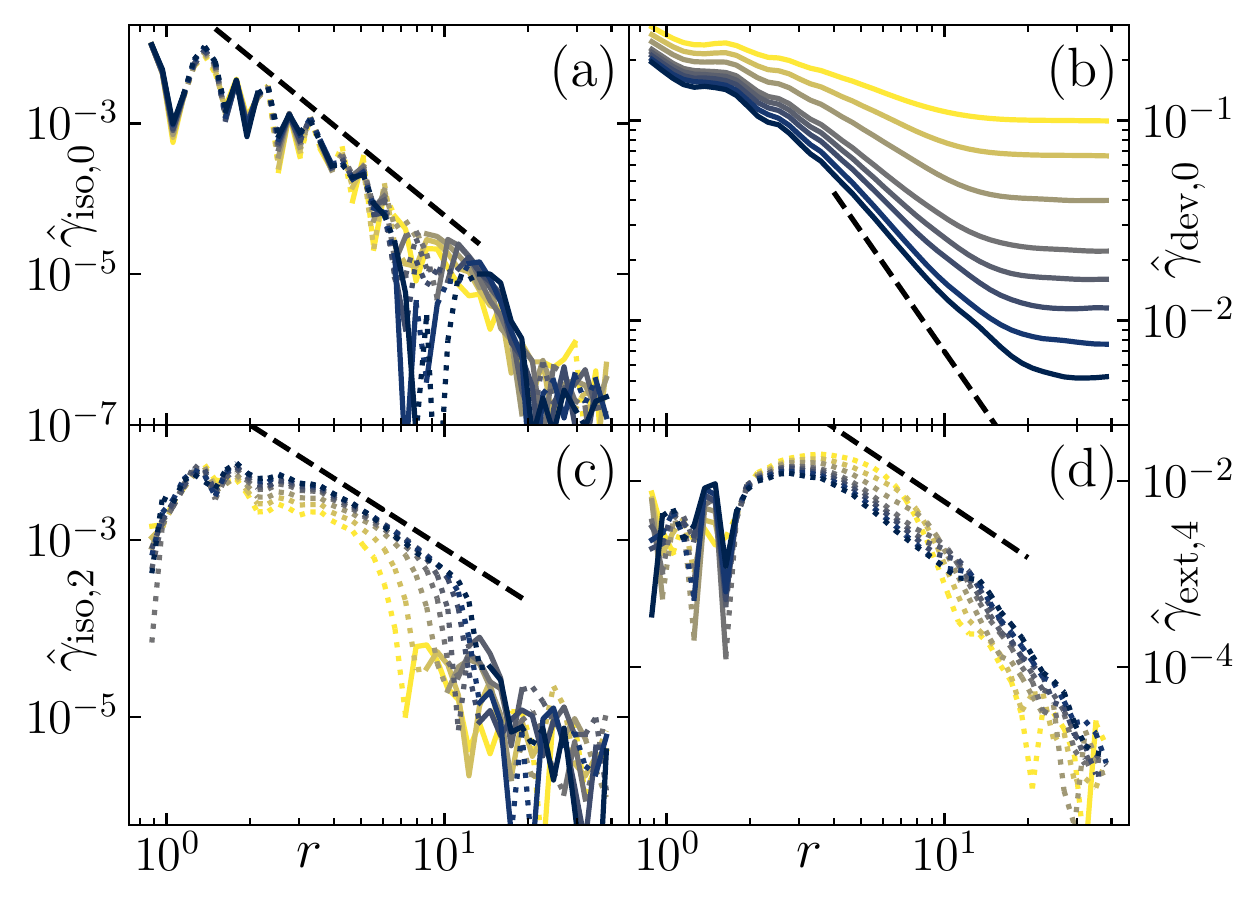}
\caption{(a-d): Fourier series cofficients $\hat{\gamma}_{\mathrm{iso},0}$, $\hat{\gamma}_{\mathrm{dev},0}$,
$\hat{\gamma}_{\mathrm{iso},2}$, and $\hat{\gamma}_{\mathrm{ext},4}$, respectively,
as a function of distance $r$ from a rearranging particle (c.f. Fig.~\ref{fig:strain_polar}).
Dotted segments denote negative values.
The colors correspond to temperatures $T = 0.100$, $0.105$, $0.110$, $0.115$, $0.120$, $0.130$, $0.140$, and $0.150$ (darkest to lightest).
Dashed lines (b-d): Slopes corresponding to the power-law decay, $\propto r^{-2}$, of an infinite elastic medium
responding to a small plastic deformation.
Dashed line (a): Power law decay $\propto r^{-3}$ plotted as an upper bound to the envelope of $\hat{\gamma}_{\mathrm{iso},0}$.
\label{fig:strain_decay}} 
\vspace{-5mm}
\end{figure}

We now assess the temperature dependence of the range of the elastic-like response by studying the angular Fourier series coefficients, which characterize the strength of the observed symmetries reported above~\cite{SM}. 
More specifically, we consider the zeroth and second order coefficients $\hat{\gamma}_{\mathrm{iso},0}$ and $\hat{\gamma}_{\mathrm{iso},2}$, the zeroth order coefficient $\hat{\gamma}_{\mathrm{dev},0}$ and the fourth order coefficient $\hat{\gamma}_{\mathrm{ext},4}$.
As shown in Fig.~\ref{fig:strain_decay},  $\hat{\gamma}_{\mathrm{iso},2}$  and $\hat{\gamma}_{\mathrm{ext},4}$ (Fig.~\ref{fig:strain_decay}(c,d))
decay for $T \leq 0.12$ as $r^{-2}$ from $r=3$ up to a cutoff that increases slightly with decreasing $T$ but is roughly $r\simeq 10$.
This power-law decay is expected for an elastic response~\cite{Zhang2020}.
The radial dependence of $\hat{\gamma}_{\mathrm{dev},0}$ (Fig.~\ref{fig:strain_decay}(b)) deviates from the elastic response behavior at higher temperatures but appears to approach the expected behavior (dashed line) with decreasing temperature. 
This can be explained by the decreasing density of rearrangements per unit time with decreasing $T$, which allows the elastic displacement field from rearrangements to extend further into the material without being obscured by other rearrangements~\cite{SM}.
Finally, for the zeroth mode of $\giso$ (Fig.~\ref{fig:strain_decay}(a)),
steady-state elasticity predicts $\hat{\gamma}_{\mathrm{iso},0} \equiv 0$.
Instead, we see a rapidly-decaying radial wave of isotropic expansion and contraction, with expansion peaks (compression troughs) roughly coinciding with the peaks (troughs) of $g \left( r \right)$~\cite{SM}.
This is consistent with the results of \cite{Zhang2020} in the context of athermal quasistatic shear: rearrangements expand the structure at the shortest distances from the rearrangement, thus locally hardening the structure, but compress the structure slightly further out, 
with this hardening-softening pattern repeating out to large distances.

\begin{figure}[t]
\includegraphics[width=0.49\textwidth]{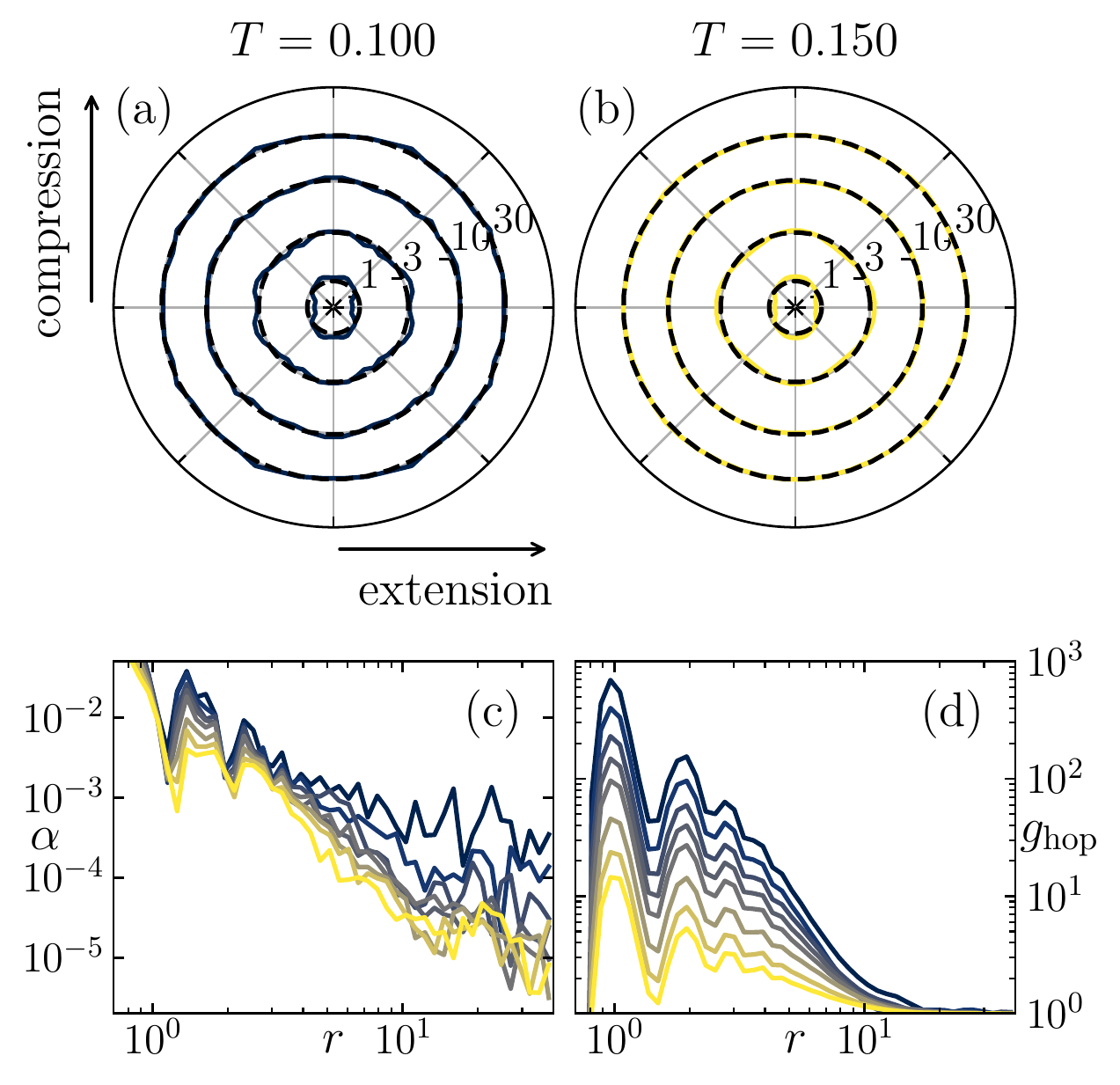}
\caption{Top row: Plots of the normalized pair distribution of rearrangements $\tghop = 2 \pi \ghop / \int_0^{2 \pi} \ghop \mathrm{d} \theta$
across one rearrangement interval $\Delta t = 10^2$ for temperatures (a) $T=0.100$ and (b) $0.150$.
Each curve corresponds to a different $r$, and is scaled so that isotropy (dashed black circle) corresponds to the
corresponding $r$ value for the log-scaled $r$ of Fig.~\ref{fig:strain_polar}.
(c): Mean squared anisotropy $\alpha(r) = \frac{1}{2 \pi} \int_0^{2 \pi} \left(\tghop - 1 \right)^2 \mathrm{d} \theta $ as a function of $r$
for $T=0.100$, $0.105$, $0.110$, $0.115$, $0.120$, $0.130$, $0.140$, and $0.150$ (lighter shade denoting higher $T$).
(d): Pair correlation of hops $\ghop(r)$, same color-coding as (c).
\label{fig:ghop_polar}} 
\vspace{-5mm}
\end{figure}

We now focus directly on dynamic facilitation and analyze its relation with the elastic responses characterized above. We use the pair distribution of rearrangements, $\ghop \left( \bm{r}\right) = \ghop \left( r, \theta \right)$
(the average density of rearrangements at $\bm{r}$ given a rearrangement at the origin,
divided by the bulk rearrangement density~\cite{SM}),
to investigate the extent to which facilitation of rearrangements is due to elasticity.
In Fig.~\ref{fig:ghop_polar}(a,b) we
display
 $\tghop(r, \theta) = 2 \pi \ghop(r, \theta) / \int_0^{2 \pi} \ghop(r,\theta) \mathrm{d} \theta$
 (the rearrangement pair distribution $\ghop \left( r, \theta \right)$ normalized at each $r$
 by its angular mean $\frac{1}{2 \pi} \int_0^{2 \pi} \ghop \left( r, \theta \right) \mathrm{d} \theta$)
 for rearrangements over the time interval $\left[ 0, \Delta t\right]$
\footnote{
By considering the same time interval for all rearrangements, we cannot tell, given a pair of rearrangements,
which rearrangement triggered the other.
However, time reversibility means that this causal ambiguity exists even for events that are well separated in time.
We have checked~\cite{SM} that choosing pairs of events in separate time intervals does not qualitatively affect our results.
}
at distances $r=1$, $3$, $10$, $30$, for temperatures $T=0.100$ and $T=0.150$.
Well below $\TMCT$, at $T=0.100$ (Fig.~\ref{fig:ghop_polar}(a)),
the spatial distribution of subsequent rearrangements remains anisotropic out to $r=30$.
The angular Fourier series expansion coefficients of $\tghop$~\cite{SM} show that
this anisotropy comes from the second- and fourth-order Fourier modes of $\tghop$,
and persists up to at least $r\approx 20$. Above $\TMCT$, at $T=0.150$ (Fig.~\ref{fig:ghop_polar}(b)), no anisostropic structure can be discerned beyond the second neighbor shell, i.e.~beyond $r>3$.

To quantify the amount of anisotropy in the spatial distribution of rearrangements we introduce the mean squared anisotropy (deviation from isotropy)~\footnote{When calculating $\alpha$ in practice, we control for effect of statistical noise,
which is larger at lower temperatures due to these temperatures having fewer rearrangements per trajectory, as described in~\cite{SM}.}
$\alpha(r) =  \frac{1}{2 \pi} \int_0^{2 \pi} \left( \tghop(r,\theta) - 1 \right)^2 \mathrm{d} \theta$, as shown in Fig.~\ref{fig:ghop_polar}(c).
High $\alpha$ at small $r$ shows that displacements at this scale, though highly nonlinear, are measurably influenced by the orientation of the linear strain. This short-ranged influence--indicative of short-ranged dynamic facilitation--persists over the entire temperature range studied.
As $T$ is lowered, the large-$r$ tail of $\alpha$ increases markedly, 
spanning two orders of magnitude, implying a significant difference in the role played by long-range elastic interactions
between the lowest and highest temperatures considered here.
At the same time, Fig.~\ref{fig:ghop_polar}(d) shows that the pair correlation for hops, $\ghop(r)$, becomes progressively longer-ranged with decreasing temperature, indicating that rearrangements are triggered further and further away.

\begin{figure}[t]
\includegraphics[width=0.48\textwidth]{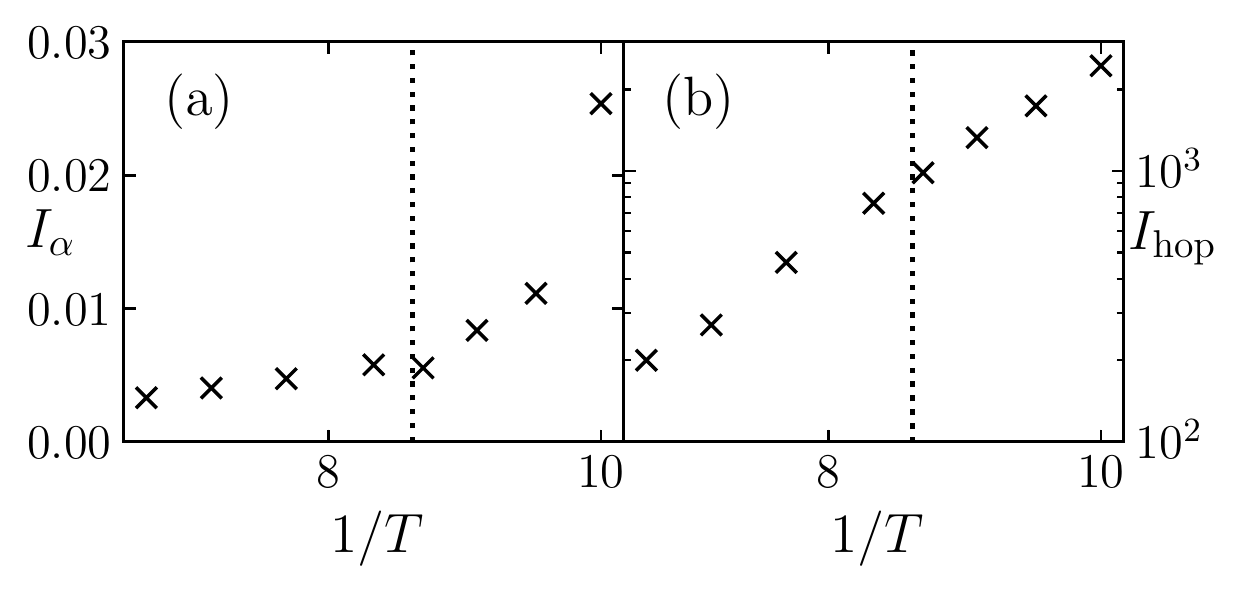}
\caption{
Integrated (a) mean squared anisotropy $I_\alpha = \int_{2.5}^{40} \alpha (r) \mathrm{d} r$ and
(b) pair distribution of rearrangements $I_\mathrm{hop} =  \int_{2.5}^{40} 2 \pi r (\ghop (r)-1) \mathrm{d} r $
as a function of inverse temperature $1/T$.
Dotted lines: $1/\TMCT$.
\label{fig:crossover}} 
\vspace{-5mm}
\end{figure}

To quantify the extent of elastically-mediated dynamic facilitation, we introduce $I_\alpha := \int_{2.5}^{40} \alpha(r) \mathrm{d} r$, which integrates $\alpha$ over distances beyond the second neighbor shell at $r \gtrsim 2.5$
(to filter out the effect of short-range facilitation)
and up to the limit beyond which finite-size effects appear.
Fig.~\ref{fig:crossover}(a) displays its evolution with $1/T$.
We observe a crossover at $\TMCT$ from a slow to a rapid growth of $I_\alpha$ as a function of $1/T$. 
Above $\TMCT$, the strain-correlated spatial organization of the rearrangements is largely obscured, beyond the second-neighbor shell, by other rearrangements~\cite{SM}.
The competition between these other rearrangements and emergent elasticity at longer distances leads to the observed crossover in $I_\alpha$.
Note that the observed crossover at $\TMCT$ is robust to choices of lower bound of the integral, rearrangement interval $\Delta t$, strain length $\xi_\mathrm{FL}$, and starting point for the integration ($r_0\geq 2.5$)~\cite{SM}.
We similarly define an integrated excess density of rearrangements (relative to the bulk rearrangement density), $I_\mathrm{hop} :=  \int_{2.5}^{40}  2 \pi r \left( \ghop(r) - 1 \right)\mathrm{d} r$, where $ \ghop(r) = \frac{1}{2 \pi} \int_0^{2\pi} \ghop(r,\theta)  \mathrm{d} \theta$.
Fig.~\ref{fig:crossover}(b) shows that $I_\mathrm{hop}$ increases with decreasing $T$, showing that hops trigger increasingly distant hops.
These results for $I_\alpha$ and $I_\mathrm{hop}$  provide direct and  explicit evidence that the elastic response to a rearrangement triggers other, distant rearrangements, demonstrating the emergence of elastic dynamic facilitation when crossing $\TMCT$.

Taken together, our findings suggest a crossover from solely short-ranged dynamic facilitation for ${T>\TMCT}$ to additional long-ranged, elastically-mediated dynamic facilitation of avalanches at ${T<\TMCT}$.
This finding provides an intriguing link between mean-field models, where $\TMCT$ marks the location where supercooled liquids first exhibit a transient solid-like behavior, and real-space dynamical heterogeneity.
We note that the crossover at $\TMCT$ is not visible in the strain field induced by a rearrangement; for example, the isotropic strain shows dipolar angular behavior at all temperatures considered.
The crossover is only evident in the spatial and angular distribution of rearrangements that follow soon after the original rearrangement at the origin.
The anisotropic spatial distribution of hops evidenced in $\ghop$, Fig.~\ref{fig:ghop_polar}(a,b), and the increase in range of $\ghop(r)$, Fig.~\ref{fig:ghop_polar}(d),
as well as the crossover at $T=\TMCT$ seen in $I_\alpha$, Fig.~\ref{fig:crossover}(a),
can only be due to the strain field created by the rearrangement at the origin.
At high $T$ strain does not facilitate rearrangements because the rearrangements are so abundant and closely spaced that their strain fields effectively interfere~\cite{SM}.
We note that at temperatures below $\TMCT$, dynamical heterogeneity still involves short-ranged facilitation, whereby the local relaxation of clusters coalesces on longer length scales to form avalanches. Nevertheless, the emergence of prominent long-ranged elastic effects must influence the spatiotemporal {\em correlations} between such avalanches.

Approximately ten years ago it was noticed that rare, long-ranged ``surges" of dynamical heterogeneity connected to strain deformation occur in supercooled liquids~\cite{WidmerCooper2009}, and that such behavior occurs in some coarse-grained kinetically-constrained models~\cite{Keys2011,Chandler2010}.
Our findings potentially pinpoint the microscopic underpinnings of such behavior, which could lead to greater understanding of the building blocks of phenomenological models of supercooled liquids.
It would be certainly worthwhile to generalize elastoplastic models \cite{Nicolas2018} to describe elastic dynamic facilitation within equilibrium thermal dynamics along the lines of \cite{popovic2020thermally,parley2020aging}. More generally, our results highlight the importance of long-ranged elastic processes in mediating the organization of dynamical heterogeneities on long length and time scales, and the need to incorporate such processes in simplified models of the glass transition.

We thank M. Ozawa and L. Berthier for providing us with equilibrated configurations and the code to generate them,
C. Scalliet for providing the swap potential files for use with LAMMPS,
and E. Corwin for sharing computational resources with us in the early stages of this project.
RNC thanks S. Ridout, G. Zhang, and I. Tah for discussions.
This work received funding from the Simons Collaboration “Cracking the glass problem” via 454935 (G. Biroli), 454945 (A. J. Liu), 454951 (D. R. Reichman) and 348126 (R. N. Chacko), and from a Simons Investigator grant (327939 to A. J. Liu).
This work used the Extreme Science and Engineering Discovery Environment (XSEDE)~\cite{XSEDE}, which is supported by National Science Foundation grant number ACI-1548562.

\bibliographystyle{apsrev4-1}
\providecommand{\noopsort}[1]{}\providecommand{\singleletter}[1]{#1}%

\end{document}